\documentstyle[prb,aps]{revtex}
\begin{document}
\twocolumn
\title{Wetting of the Au(110) substrate: Homoepitaxial islands
and layers}

\author{ G. Bilalbegovi\'c }

\address{
Department of Physics, Institute ``Rudjer Bo\v skovi\'c'',
P.O.Box 1016, 10001 Zagreb, Croatia}

\maketitle

\begin{abstract}

The morphology of epitaxial structures formed in solidification of
liquid Au films on the
Au(110) surface was studied by molecular dynamics
simulation based on a many--body interatomic  potential.
The $(1\times 2)$ reconstructed and smooth phase at
the temperature $T=500$ K, as well as
a deconstructed and rough phase of the Au(110) substrate at $T=900$ K,
were investigated.
The three--dimensional islands with the $(111)$ oriented facets
were formed in  solidification of thinner liquid films.
At the same time, the substrate below these islands underwent a weak
faceting process. Conversely, the solidification of thicker
liquid films resulted in the flat solid films.
The two surface phases of the Au(110) substrate induced
different structure of these epitaxial solid films. The films
were studied via construction of the Voronoi polyhedra,
density profiles, and surface stress calculations.
\end{abstract}



\section{Introduction}

The wetting of substrates by liquid and solid adsorbates
attracts interest in diverse fields. The connection of
wetting and epitaxial growth is important for fundamental
surface physics and at the same time for applications.
The Au(110) surface has been investigated by many
experimental techniques \cite{Campuzano,Marco}.
In a temperature range from $T=0$ K up to the bulk melting point
$T=1337$ K,
the Au(110) surface exhibits several structural phases.
The surface is at low temperatures smooth and
reconstructed into the so--called ``missing--row''
$(1\times2)$ structure. In this phase the length of
the surface unit
cell is the same as in the bulk along $[1\bar{1}0]$,
while it is twice as long along the $[ 001 ]$ direction.
Alternating $[1\bar{1}0]$ rows are absent in the topmost layer.
The missing row structure disappears  in a phase transition
of the Ising type at $\approx 650$ K.
In recent experiments the roughening transition on some
well--characterized surfaces was studied \cite {Lapujoulade}.
At the roughening transition temperature
the flat facet on equilibrium crystal shape becomes rounded and
the step free energy of an interface goes to zero.
For statistical mechanical models it is
suitable to define the roughening transition where
the height--height correlation function exhibits divergence.
In particular, the Au(110) surface undergoes the  Kosterlitz--Thouless
roughening transition at $\approx 700$ K \cite{Marco,Lapujoulade}.
The Au(110) surface also exhibits surface melting:
it starts to melt below the bulk melting temperature $T_m$
and the thickness of the liquid layer diverges as  $T_m$ is approached
\cite{Hoss}.

Barbier, Salanon, and  Spr{\" o}sser recently started investigation
of the structure and ordering dynamics for the growth of Au
on the $(1\times 2)$ reconstructed
Au(110) surface \cite{Barbier}.
They have found that a metastable $(1\times 1)$
phase is formed after deposition of a half of a monolayer at $190$ K.
The $(1\times 2)$ disordered phase is formed after a deposition
of one monolayer.
The method of deposition of Barbier and coworkers included
some features of the
epitaxy from the liquid phase. They used a Knudsen type evaporator for
liquid gold and obtained a flux of one monolayer in a few minutes.
The presence of several surface phases on Au(110)
calls for a detailed investigation of homoepitaxial growth on
this substrate.
Unfortunately it is often a formidable task to study solidification
and growth at
an atomistic level using now available experimental techniques.
Computer simulations employing carefully chosen potentials for real
materials may certainly help one to understand these phenomena.
For example, results of
MD simulations with the Stillinger--Weber potential are in good
agreement with experimental observations of
the silicon surface--melt interface after laser-induced melting
\cite{Landman,Abraham}. MD simulation study of a  metal
surface,  in particular solidification of thin liquid Au films
on Au(110), is presented in the following.

The structures formed in solidification of liquid Au films
on the $(1\times2)$ reconstructed phase, as well as on
a  rough phase of the Au(110) surface are studied in this work.
The MD simulation method is used to investigate the microscopic
features of epitaxy.
The results show that three-dimensional islands are formed at
the substrate in  solidification of thinner liquid films,
whereas flat solid films grow in solidification of thicker films.
Different structures of epitaxial films are found for a reconstructed
and rough substrate. The structure of the films was investigated via
monitoring of real--space particle trajectories,
density profiles, surface stress
values, and construction of the Voronoi polyhedra.
In the following the MD technique, and the potential and simulation
methods are described in Sec. II.
Results are presented in Sections III--V. A discussion and the conclusions
are given in Sec. VI.

\section{Simulation Model and Method}

MD simulations were used  to study epitaxial solidification
of liquid Au films on Au(110).
The interatomic interactions were derived from a many--body
classical Hamiltonian \cite{Daw,Nieminen,Furio}.
This kind of potentials gives a good characterization of interactions
and cohesion in metals \cite{Nieminen}.
All details of the so--called ``glue'' model
potential were fixed before
and there were no arbitrary parameters in the present simulation
\cite{Furio}. This potential is given by
\begin{equation}
V=\frac{1}{2} \sum_{i,j=1 \atop (j\neq i)}^N \phi(r_{i,j}) +
\sum_{i=1}^N U(n_i),
\label{eq:1}
\end{equation}
where the density $n_i$ is a superposition of two--body contributions
\begin{equation}
n_i= \sum_{j=1 \atop (j\neq i)}^N \rho (r_{i,j}).
\label{eq:2}
\end{equation}
In the glue model all three functions $\phi (r)$, $U(n)$ and $\rho (r)$
were included in the fitting procedure \cite{Furio}. The glue potential
for gold is well tested and known to reproduce experimental results
for many diverse bulk, surface and cluster properties.
Garofalo, Ercolessi, and Tosatti used this potential
to study the reconstruction on the Au(110) surface \cite{Mario}.
They proved that the $(1\times 2)$ missing
row structure has the lowest surface energy.
Bernasconi and Tosatti
studied the Au(110) surface structure within the glue model
at temperatures around deconstruction and roughening \cite{Marco}.
These transitions in MD simulation are characterized with
slow kinetics. In addition, a large lateral
size of MD box is necessary due to
the logarithmic divergence of the correlation
function at the roughening transition. Therefore, Bernasconi and Tosatti
showed that it is not
feasible nowadays to study the thermodynamical characteristics of
the roughening transition on Au(110) using MD method.

The simulation started from  the model for the substrate,
but with a goal of modeling the epitaxy on Au(110).
Separate MD slabs were used to study the growth on
a reconstructed and rough surface.
The temperature was controlled by rescaling the particle
velocities. The lattice constants of all samples were changed
with temperature according to the expansion coefficient of the bulk.
This coefficient and the bulk melting temperature ($T_m\approx 1355$ K)
were known from the previous bulk simulation \cite{Furio}.
The melting point in the glue model is in good agreement with
the experimental value of $1337$ K.
MD boxes consisted of 12 layers and were constructed to have the
$(1\times 2)$ reconstructed top layer at $T=0$ K.
The three bottom layers were kept fixed to model the bulk.
The number of
atoms was 10 along the $[001]$ direction and 5 along
$[1\bar{1}0]$. These periodically repeated MD slabs
for the substrate were made up of 575 particles.
The simulations for MD boxes with 10 and 20 atoms along
the $[1\bar{1}0]$ direction (consisting of
750 and 1500 particles, respectively) were also performed.
No size dependence was observed, and only the detailed
results for the smallest MD box will be presented here.
As usual, for bigger samples and larger phase
space longer MD simulations are needed to achieve the same structure.
This requirement is crucial for MD studies of thin film growth.
The samples were heated  up to $T=500$ K $= 0.37 T_m$
for a reconstructed phase,
and $T=900$ K $=0.66 T_m$ for a rough phase.
At these temperatures runs of $5 \times 10^5$ MD time steps
($3.6$ ns) for a smooth phase and $10^6$ MD steps ($7.1$ ns)
for a rough phase were done. Bernasconi and Tosatti found that
a long equilibration is necessary
to obtain the rough phase on Au(110) \cite{Marco}.
The final configurations of MD boxes representing the Au(110) substrate
are shown in Fig.~\ref{fig1}. Two and four additional solid layers
were then deposited on the top of MD boxes to follow the stacking of
the Au(110) surface. In this paper these new epitaxial atoms are
referred to as thinner films (two initial solid layers) and
thicker films (four initial solid layers).
The structures of the Au(110) substrate
were preserved by temporarily fixing the corresponding particles.
The epitaxial films were warmed up to $T=2000$ K $=1.48T_m$,
and then equilibrated for $0.36$ ns.
No  atom was observed to evaporate from the slab surface.
This agrees with the experimental evaporation rate of gold
which is in the range from $50$ atoms s$^{-1}${\AA}$^{-2}$ at 1360 K to
$6\times 10^4$ atoms s$^{-1}${\AA}$^{-2}$ at $3000$ K \cite{Handbook}.
The samples with liquid film on the top
were cooled back to $500$ K and $900$ K
and the particles representing the substrate were
released. The time evolution of substrates and epitaxial films was
then followed for additional $3.6$ ns. Some simulations for deposition
and
solidification were repeated for the second MD box of the same size and
similar results for the morphology of epitaxial structures were obtained.

\section{Morphology}

Figure \ref{fig2}(a) shows the  result of solidification
of a thinner
liquid film on the smooth, $(1\times 2)$ reconstructed substrate after
$5\times 10^5$ time steps. The faceted
structure of the solid film is evident.
The big island consists of two connected
four--layer high islands. The facets are with the $(111)$ structure and
they are oriented under the angle of $35 ^\circ$ to the substrate. A small
island on the left was also formed.
Figure \ref{fig2}(b) shows the result of solidification
of a thinner liquid film on the rough Au(110) surface
after $3.6$ ns of simulation.
One four-layer high island with a flat top was formed.
The facets on the island are disordered,
with the $(111)$ structure and oriented under the angle of $35^\circ$
to the substrate. Underlying substrate exhibits faceting: four substrate
atoms (two on each side of the box) moved to the island and as a result
longer $(111)$ facets were formed.
The $(1\times2)$ reconstructed substrate shown
in Fig.~\ref{fig2}(a) also tends to facet after
solidification of a liquid film.

Figure \ref{fig3} represents the structure formed in
solidification of a thicker liquid film
on the smooth, $(1\times2)$ reconstructed Au(110)
surface after $3.6$ ns of
simulation. The film is flat and its surface is disordered.
Deposited atoms first filled the missing rows of the substrate and then
more disordered layers were formed on the top of the film. The observed
trends
agree with the results of Barbier {\it et al} \cite{Barbier}.
The structure formed in solidification of thicker liquid film on
a deconstructed, rough substrate after $3.6$ ns of simulation is shown
in Fig.~\ref{fig4}. The film is flat and more ordered
along the $[100]$ direction than along
$[1\bar{1}0]$. The side view in Fig.~\ref{fig4}(a) shows that stacking
faults exist at the rough substrate--film interface.
The top layers of the rough substrate are more ordered after than
before deposition. This is the case for both thinner and thicker films
[shown in Figs.~\ref{fig2}(b) and 4, respectively].

\section{Structure}

The reason for the island formation in deposition of thinner films was
a tendency of gold atoms to increase their coordination. It is well known
that a small coordination of atoms at gold surfaces provides
a driving
force for their reconstruction, i.e., the formation of denser layers with
a higher coordination. It was calculated that the average coordination was
$6.044$ neighbors for the initial, ideal solid configuration
of a thinner film. The average coordination increased to
9.068 neighbors for the initial solid configuration of a thicker film.
The small coordination of $\approx 6$ caused the  clustering,
i.e., the change of the structure that increased the coordination number.
The three--dimensional islands, therefore,  were formed in solidification of
thinner films. The higher coordination of atoms in thicker epitaxial films
was of the same order as for reconstructed gold surfaces and the
flat films were formed.

The densities projected on the vertical axis for the flat
epitaxial films are shown in Fig.~\ref{fig5}.
The films are less ordered than the substrate, but the
layered structure exists on the density plots.

Construction of the Voronoi polyhedra was done to describe
an amorphous structure of the flat solid films \cite{Collins}.
The Voronoi polyhedron is a
topological generalization of the Wigner--Seitz cell for a crystal.
It is defined for a given atom as the region consisting of all
points nearer to it than to any other atom. The Voronoi polyhedra
characterize the local atomic configurations and disorder
for a given phase.
Figure \ref{fig6} shows the distribution of the number of neighbors
for the Voronoi polyhedra constructed for
the atoms in the flat solid films.
The coordinates of particles are taken from
MD simulations after the time evolution of $3.6$ ns, i.e.,
for the structures shown in Figs.~\ref{fig3} and \ref{fig4}.
Only the particles in the films were analyzed,
but the atoms in the first layer of the substrate were also
considered as
neighbors. The particles were not assumed to be neighbors
beyond the cutoff of $5$ {\AA}.
The average number of the Voronoi neighbors for a thicker film
on a smooth substrate is $<i>=\sum_i in_i = 14.19$,
whereas for a thicker film on a rough substrate $<i>$ is $14.96$.
These values agree with other
reported results for the number of faces of the Voronoi polyhedra in
three--dimensional systems \cite{Collins}.
The average value of this number is
$6$ in two dimensions, whereas it is $14$
in three dimensions and for regular close packing.
Using MD simulation Rahman
found $15.67$ for the randomly distributed atoms of ideal gas, $14.45$
for the liquid argon and $14.26$ for the solid argon \cite{Collins}.
Finney found the value of $14.2$ for a random close packing of hard
spheres using the Monte Carlo simulation \cite{Collins}.
Figure \ref{fig6} shows that the maximum of distribution for
the number of faces of the Voronoi polyhedra corresponds to $i=13$
for the film on a smooth substrate,
and to $12$ for the film on a rough substrate. The distributions are
broad and represent a disorder, i.e.,
the presence of topologically
defective polygons where the number of faces is not $14$.

\section{Surface stress}

When the film is deposited on the surface at certain temperature there is
a competition between the requirements of bonding to the substrate and
a tendency of particles to adopt the minimum energy separations in
the epitaxial layers. These interactions are origin of the surface stress
in the film. The surface of the substrate is by itself in a stressed state
because of different bonding than in the bulk.
The surface stress
tensor $\sigma_{\alpha\beta}$ is given by
\begin{equation}
\sigma_{\alpha\beta}=\gamma \delta_{\alpha\beta} + \frac{\partial\gamma}
{\partial\epsilon_{\alpha\beta}},
\label{eq:3}
\end{equation}
where $\gamma$ is the surface tension, $\delta_{\alpha\beta}$ is
the Kronecker symbol, $\epsilon_{\alpha\beta}$ is the surface
strain tensor, and $\alpha$, $\beta$ are directions in the surface plane
\cite{Cammarata}.
In MD simulation the surface stress can be obtained from the components
of the pressure tensor \cite{Broughton}. In computation of the stress MD
slabs for the substrates and thicker films described in Sec. II were
used. MD box representing the $(1\times 2)$ reconstructed Au(110) surface
at $T=0$ K was equilibrated for $0.71$ ns.
The values of surface stress along the $[1{\bar 1}0]$
and $[001]$ directions that describe the anisotropy of the $(1\times 2)$
reconstructed Au(110) substrate are presented in Table \ref{table1}.
The components of
the surface stress tensor along and perpendicular to the close packed rows
of a reconstructed substrate are different even for the films.
The stress is smaller along the close--packed rows for
a reconstructed Au(110) surface at $T=0$ K.
As the temperature increases this value becomes bigger than the other
component, for both the substrates and films. The same relationship was
obtained by  Toh, Ong and Ercolessi for the Pb(110) surface \cite{Toh}.
Table \ref{table1} also shows that the stress for the $(1\times 2)$
Au(110) surface
at $500$ K decreases after deposition of the film. On the contrary,
the surface stress increases when the film is deposited on the rough
substrate at $900$ K.
In general, the surface stress values found in these simulations
are similar to those found in other recently reported calculations for
the $(110)$ surfaces in  fcc metals
\cite{Toh,Needs,Varga,Wolf,Feibelman}. The surface
stress can be in principle calculated by ab--initio electronic
structure calculation and MD simulation.
For all $(110)$ surfaces the stress is positive, i.e.,
tensile and favors contraction of the lattice.
The experimental data for the surface stress on well--defined
metallic surfaces do not exist.
The surface stress value of $0.073$  eV{\AA}$^{-2}$
was obtained at $323$ K in measurements for the radial strain
in small Au spheres \cite{Cammarata}.
Solliard and Flueli found (at $300$ K) the value
$0.1922$ eV{\AA}$^{-2}$ for
the $(220)$ ring on a small gold sphere, and $0.1991$ eV{\AA}$^{-2}$
for the $(422)$ ring  \cite{Solliard}.
For comparison, the average values of the surface stress tensor
for the substrates and films studied
in this work are also given in Table \ref{table1}. More studies, and
especially measurements, for the stress on metal surfaces and films
are necessary. The change in the surface stress after deposition of
a layer of Ga on the Si crystal was measured \cite{Martinez}.

\section{Discussion and Conclusions}

The results of simulation show  solidification of liquid Au
films on the Au(110) surface at $500$ K and $900$ K.
This problem is equivalent to the wetting of the Au(110)
substrate by  thin films of gold \cite{Schick}.
The morphological change of the film
with its thickness is a transition from nonwetting to wetting behavior.
The initial deposited solid films were warmed up to $T=2000$ K.
Therefore undercoolings of $1500$ K for a smooth
phase and $1100$ K for a rough phase of the substrate were used in
the simulations.
At these large undercoolings the growth and solidification can be viewed
as continuous processes, i.e., a nucleation rate is high \cite{Krumbhaar}.
As a consequence, at both temperatures
solidification may proceed from any point on the surface and there is no
requirement for the special nucleus as in a nucleation mechanism. Therefore
in this simulation no difference in the growth mode between a smooth and
rough substrate was found. This is consistent with
a view that in solidification smooth surfaces undergo
a kinetic roughening transition already for small driving forces
\cite{Krumbhaar}. As a result only rough surfaces are
effectively present in solidification. The $(1\times 2)$ reconstruction
on the smooth substrate was naturally removed because the deposited atoms
first filled the missing rows.

The facets on the three--dimensional islands formed in solidification
of thinner films were $(111)$ oriented.
The substrate below these islands exhibited faceting. The small facets
of the substrate joined the big $(111)$ facets of the islands.
The $(111)$ facets are also
formed in faceting of gold vicinals induced by surface reconstruction
and surface melting \cite{Goranka}. The $(111)$ face of fcc metals has
the lowest surface energy and the structures develop to ``open'' the
$(111)$ orientation. Recently faceting induced by deposition of
metallic films was found for the bcc substrates, such as
W(111) and Mo(111) \cite{Madey}.
For example, a monolayer of Pd on W(111) causes the  formation
of the $(211)$ oriented facets. Although faceting is already
in the focus of the
current interests in surface physics, this phenomenon certainly needs
further investigation \cite{Williams}. This is especially true for
faceting of the substrates induced by deposition of thin films.

The method of the film preparation used in this work resembles
the high--temperature annealing in simulations for
a structural investigation of free clusters. The techniques of thermal
annealing and rapid thermal annealing are also used in experiments for
crystallization of amorphous thin films.
The direct high--temperature annealing procedure
gives (within reasonable MD simulation time) amorphous structures
of low energies for clusters bigger than $60$ particles
\cite{Wanda}. In this simulation the substrate
tends to order the particles during solidification. Similar
method of the high--temperature surface annealing of free Au clusters
gives better results than other techniques \cite{Wanda}. In this method
only the particles in the external shell are allowed to move, while the
internal particles of the cluster are kept frozen in the crystalline
structure. The method of the film preparation at the  high temperature,
used in this work, simulates the deposition from the liquid phase and
should
be important for MD studies of epitaxy. It is well known that in MD
simulations of the film growth from the vapor phase a big problem is
insufficient computer time to follow experimental deposition rates.
MD simulations done with extremely high deposition rates sometimes produce
unphysical effects. The simulations where the solid films
of a certain thickness are directly placed on the substrate are far from
the experimental reality. In addition,
if such a  method is used at low temperatures
and for thick films, then it is difficult
to achieve real equilibrium structure in a limited simulation time.
The method of the high--temperature annealing of thin films
gives different solution for MD simulation of epitaxy.

In summary, solidification of liquid Au films on the
Au(110) substrate was studied.
MD simulation method, based on a many--body
interatomic potential of proven accuracy, was used.
The results showed that three-dimensional islands were formed
in solidification of thinner liquid films.
The substrate below these islands exhibited faceting.
Flat solid films grew in homoepitaxial solidification
of thicker liquid films. The films deposited
on the smooth, $(1\times 2)$ reconstructed phase at $T=500$ K
and on the rough, deconstructed
phase of the substrate at $T=900$ K were studied. Epitaxial solid films
of different structure were formed on these substrates.  Surface phases
of Au(110) did not change the wetting mode of homoepitaxial films.
Instead, the thickness of the film
determined homoepitaxial growth mode on the Au(110) substrate.
The detailed experimental
studies of homoepitaxial growth on the Au(110) substrate,
at all temperatures and for a various thicknesses of the film,
are necessary.

\acknowledgments

I would like to thank F. Ercolessi, M. Milun, P. Pervan, and
E. Tosatti for discussion and help.

\clearpage

\begin{figure}
\caption{
Atomic configurations of molecular dynamics slabs used as
a substrate. Two boxes along the [$1{\bar 1}0$] direction are shown.
(a) The $(1\times 2)$ reconstructed and smooth surface: a missing row
structure is present at $500$ K.
(b) A deconstructed and rough surface at $900$ K:
the missing row structure of the substrate disappears.}
\label{fig1}
\end{figure}

\begin{figure}
\caption{
Particle trajectories showing the side view of molecular dynamics boxes
after an equilibration time  of $3.6$ ns for a thinner film:
(a) on the $(1\times 2)$  reconstructed Au(110) surface at $500$ K,
(b) on a rough Au(110) surface at 900 K.
All trajectory plots refer to a time span of $7$ ps and
show only the moving particles. For the side views trajectories display
only nine substrate layers and the five [$1{\bar 1}0$] planes overlapped.
The arrows point to the top substrate layer.
Note faceting of the substrate in Fig. 2(b): four atoms in the two top
layers of the substrate are missing, remaining microfacets are $(111)$
oriented and join the facets of the island.}
\label{fig2}
\end{figure}

\begin{figure}
\caption{
Thicker film on the $(1\times 2)$ reconstructed Au(110) substrate
at $500$ K.
(a) Side view of the box. Deposited particles first fill the missing rows
in the top (i.e., ninth on this figure) layer of the substrate, the next
layer of the film is regular and follows the stacking of the substrate.
The top region of the film is amorphous.
(b) Top view of the film.}
\label{fig3}
\end{figure}

\begin{figure}
\caption{
Thicker film on a rough Au(110) surface at 900 K.
(a) Side view of the box along the $[100]$ direction: the top
(i.e., ninth on this figure) layer of the substrate
is more ordered below the film then before deposition
[corresponding rough substrate is shown in Fig. 1(b)].
(b) Top view of the film.
(c) Side view of the box along the $[1\bar{1}0]$ direction.}
\label{fig4}
\end{figure}

\begin{figure}
\caption{
The $(x,y)$ averaged density profile for
molecular dynamics box simulating solid films grown from thicker
liquid films. The three leftmost layers of the box are kept fixed
and the film is on the right.
(a) Film on a reconstructed substrate at $500$ K.
(b) Film on a rough substrate at $900$K.}
\label{fig5}
\end{figure}

\begin{figure}
\caption{
The fractional concentrations for the number of neighbors, i.e.,
faces of the Voronoi polyhedra.
(a) Thicker film on the $(1\times 2)$ reconstructed Au(110) substrate
at $500$ K.
(b) Thicker film on a rough Au(110) substrate at $900$ K.
The numbers of neighbors not equal $14$ represent the extent
of amorphousness.
The thick vertical lines show the numbers, whereas the thin line is
the envelope.}
\label{fig6}
\end{figure}

\clearpage

\begin{table}
\caption{
Surface stress $\sigma_{\alpha\beta}$. The components of the
stress along and perpendicular to the close packed rows
of the $(1\times 2)$
reconstructed Au(110) surface, as well as their average value are shown.
All values are in eV{\AA}$^{-2}$.}
\label{table1}
\begin{tabular}{l l l l}
Surface structure & $[1{\bar 1}0]$
 & $[001]$  & Average \\
\hline
Film on the smooth substrate, $500$ K & $0.101$ & $0.063$ & $0.082$ \\

Film on the rough substrate, $900$ K & $0.155$ & $0.124$ & $0.140$ \\

$(1\times 2)$ Au(110), $500$ K & $0.199$ & $0.148$ & $0.174$ \\

Rough Au(110), $900$ K & $0.130$ & $0.118$ & $0.124$ \\

$(1\times 2)$ Au(110), relaxed, $0$ K & $0.184$ & $0.191$ & $0.188$ \\
\end{tabular}
\end{table}

\end{document}